\documentclass[letterpaper, 10.5 pt, conference]{ieeeconf}  

\IEEEoverridecommandlockouts                              
\usepackage{graphicx} 
\usepackage{bm}
\usepackage[utf8]{inputenc}
\DeclareUnicodeCharacter{202F}{\,}

\usepackage{xcolor}
\usepackage{hyperref} 
\usepackage{comment}

\begin{document}
\title{\LARGE \bf
Power Stabilization for AI Training Datacenters
}

\author{
Esha Choukse, Brijesh Warrier, Scot Heath, Luz Belmont, April Zhao, Hassan Ali Khan, Brian Harry$^1$, \\
Matthew Kappel, Russell J. Hewett$^1$, Kushal Datta, Yu Pei, Caroline Lichtenberger, John Siegler,  \\
David Lukofsky$^1$, Zaid Kahn$^1$, Gurpreet Sahota, Andy Sullivan, Charles Frederick,  Hien Thai, \\
Rebecca Naughton$^1$, Daniel Jurnove, Justin Harp$^1$, Reid Carper, Nithish Mahalingam, \\
Srini Varkala, Alok Gautam Kumbhare, Satyajit Desai,  Venkatesh Ramamurthy, \\
Praneeth Gottumukkala, Girish Bhatia, Kelsey Wildstone, Laurentiu Olariu, \\
Ileana Incorvaia, Alex Wetmore, Prabhat Ram, Melur Raghuraman\\
Mohammed Ayna, Mike Kendrick, Ricardo Bianchini \\
\textbf{Microsoft} \\ 
\\
Aaron Hurst, Reza Zamani, Xin Li, Michael Petrov, Gene Oden, Rory Carmichael \\
 \textbf{OpenAI} \\ 
\\
Tom Li, Apoorv Gupta, Pratikkumar Patel, Nilesh Dattani, Lawrence Marwong, Rob Nertney, \\
Hirofumi Kobayashi, Jeff Liott, Miro Enev,
Divya Ramakrishnan, Ian Buck, Jonah Alben\\
 \textbf{NVIDIA}\\
}

\maketitle
\thispagestyle{empty}
\pagestyle{empty}

\begin{abstract}

Large Artificial Intelligence (AI) training workloads spanning several tens of thousands of GPUs present unique power management challenges. These arise due to the high variability in power consumption during the training. Given the synchronous nature of these jobs, during every iteration there is a computation-heavy phase, where each GPU works on the local data, and a communication-heavy phase where all the GPUs synchronize on the data. Because compute-heavy phases require much more power than communication phases, large power swings occur. The amplitude of these power swings is ever increasing with the increase in the size of training jobs. An even bigger challenge arises from the frequency spectrum of these power swings which, if harmonized with critical frequencies of utilities, can cause physical damage to the power grid infrastructure. Therefore, to continue scaling AI training workloads safely, we need to stabilize the power of such workloads. This paper introduces the challenge with production data and explores innovative solutions across the stack: software, GPU hardware, and datacenter infrastructure. We present the pros and cons of each of these approaches and finally present a multi-pronged approach to solving the challenge. The proposed solutions are rigorously tested using a combination of real hardware and Microsoft's in-house cloud power simulator, providing critical insights into the efficacy of these interventions under real-world conditions.

\end{abstract}

\footnotetext[1]{Work was done when they were employees at Microsoft.}

\section{INTRODUCTION}

The past few years have seen a fast increase in the size and speed of deployment of training clusters for frontier foundational large AI models~\cite{Supermicro_xAI_Colossus_2024,pilz2025trendsaisupercomputers}.
A single training job can span more than a hundred thousand GPUs, ~\cite{OpenAI_ScalingK8s_2021, grok1_xai_2024}.
During a training job, the participating GPUs work in lockstep under the bulk synchronous paradigm~\cite{OpenAI_TechniquesTraining_2022,shoeybi2019megatronlm}.
Although most of the training time is spent doing computational work, there are phases where all the participating GPUs need to communicate, or do other tasks.
This could be during the all-reduce communication collective in an iteration~\cite{OpenAI_TechniquesTraining_2022}, where the participating GPUs need to synchronize on the model-weight update, or when a checkpoint is being recorded.
While the power utilized in a computationally heavy phase in a GPU can be close to its TDP (Thermal Design Power), the power utilized during the communication phase can be close to the idle power.

\begin{figure*}[tbh]
    \centering
    \includegraphics[width=0.9\linewidth]{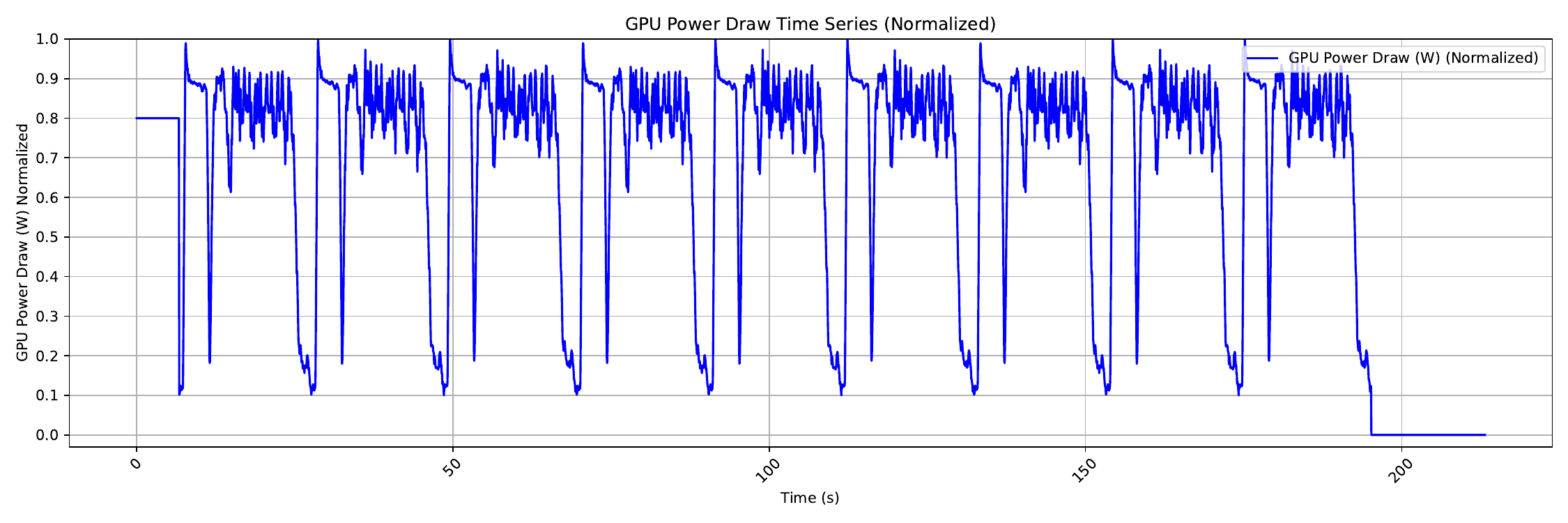}
    \caption{Power readings from an at-scale training job on DGX-H100 racks.}
    \label{fig:powerbg}
\end{figure*}

Such huge variations in power draw lead to swings at the node level (Figure~\ref{fig:powerbg}). Furthermore, due to the large and synchronous nature of the job, participating nodes are co-located to form a majority of a datacenter, or even multiple datacenters in the same grid, making the power swings visible at the rack, datacenter, and power grid levels.
At scale, these swings can amount to tens or hundreds of megawatts, occurring at frequencies that, if poorly aligned with the resonant characteristics of power grid components (e.g., turbine generators or long transmission lines), can risk grid instability and mechanical failure. These issues are not theoretical — multiple utility providers have now documented the impact of harmonics induced by synchronized computing loads~\cite{nerc2019disturbance}.

This paper proposes a multi-faceted approach to mitigate these risks, grounded in real-world production telemetry. We first characterize the amplitude and frequency of power oscillations observed in hyperscale training clusters. Then, we evaluate three classes of mitigation: (1) software-based approaches that inject controlled workloads to smooth transitions, (2) GPU-level firmware features that enforce ramping constraints and power floors, and (3) rack-level energy storage to absorb and release power as needed. Each technique is assessed for its effectiveness, energy efficiency, and deployability. Finally, we advocate for cross-industry co-design across software, hardware, and infrastructure to ensure that AI systems remain both scalable and power-aware.

\section{Background and Motivation}

\subsection{Rise of large training jobs}
Over the past several years, the field of artificial intelligence has seen explosive growth in the size and complexity of neural networks. Originally, popular models such as AlexNet~\cite{krizhevsky2012imagenet} contained on the order of 60 million parameters and could be trained on a single GPU. However, in the time since, the community has rapidly embraced “foundation models” with billions, and even hundreds of billions of parameters.
The scaling of AI models has been driven by reinforcing trends in: (1) architectural advances such as Transformers, (2) system design strategies like data, tensor, and pipeline parallelism that enable the distribution of computation, (3) hardware evolution, including higher bandwidth interconnects and specialized datacenter GPU devices, and (4) datacenter deployment sizes that have grown from $<10 MW$ to $>100 MW$.

In particular, the training of GPT-3 (175B parameters)~\cite{brown2020languagemodelsfewshotlearners} and successors like Grok1 (314B parameters)~\cite{grok1_xai_2024}, PaLM (540B parameters)~\cite{chowdhery2022palm}, Llama3.1 (405B parameters)~\cite{touvron_llama_2023}, etc., has showcased the trend toward massive-scale models that demand equally massive computational resources. While domain-targeted training, model size efficiency, and the use of pretrained backbones has recently enabled cheaper training runs for model families like Phi~\cite{abdin2024phi3} and DeepSeek~\cite{deepseek2024v3}, trends show that the foundation model training will still span several tens-of-thousands of GPUs.

\subsection{Compute and communication phases}
Large-scale training of models proceeds in iterations, each iteration going over a small batch of the training data, encompassing distinct \textit{compute and communication phases}. The model and the data are distributed across the participating GPUs using various parallelism techniques~\cite{shoeybi2019megatronlm}. In the forward pass of each iteration, each GPU processes its own portion of the mini-batch independently, calculating partial predictions based on the current model parameters. Subsequently, during the backward pass, these devices compute local gradients of the loss function with respect to their assigned subset of data. Because all participating GPUs must converge to the same global model state, the locally computed gradients must be aggregated and shared after each iteration.

This aggregation is typically performed via an all-reduce operation, most commonly implemented using optimized libraries (e.g., NVIDIA NCCL~\cite{nccl2025}). Through the all-reduce, partial gradients are summed or averaged across all GPUs, ensuring that every model replica obtains an identical final gradient vector. After this communication step, model parameters are synchronously updated on each GPU, thereby preserving consistency across the distributed system. This iteration of local computation and all-reduce communication forms the core of bulk-synchronous parallelism~\cite{OpenAI_TechniquesTraining_2022}, wherein training cannot advance to the next iteration until all devices complete the gradient synchronization phase.

Less frequent communication phases occur during checkpointing, when model states are periodically saved to persistent storage. Checkpointing preserves the training state at regular intervals, safeguarding long-running jobs against common hardware or software failures~\cite{meta2025aireliability}. By periodically writing model parameters—and, in some cases, optimizer states—to persistent storage, checkpointing enables a faster and more consistent recovery, allowing training to resume without losing significant computational effort or compromising convergence guarantees.Though not as frequent as gradient synchronization, checkpointing can still introduce non-trivial network and I/O overhead across the cluster.

Although techniques for overlapping communication and computation (e.g., dedicated DMA engines or asynchronous kernel launches) can reduce idle time, most data-parallel workloads retain a significant synchronization step at the end of each iteration~\cite{shoeybi2019megatronlm,deepseek2024v3}. Checkpointing and other transient phases, such as failure recovery or dynamic load balancing, can further exacerbate power variations by introducing additional patterns of activity and inactivity across the GPU fleet.
Asynchronous training methods with lazy weight updates avoid this synchronicity challenge, but have been shown to have accuracy and convergence trade-offs~\cite{lian2018asynchronous}. 

\subsection{Power consumption during training}
The power draw of GPUs can rapidly swing as the application transitions between compute and communication phases (Figure~\ref{fig:powerbg}). In the compute phase, the GPU tensor cores typically run at or near full utilization, drawing power close to the TDP of the device~\cite{patel2024llmpower}. By contrast, during the communication phase, in which the GPUs synchronize gradients via collective operations (e.g., all-reduce), or are working on checkpointing the model, compute resources sit idle or underutilized. As a result, the same GPU may exhibit a dramatic drop in power consumption over relatively short timescales — ranging from once per second or less, to once every tens of seconds, depending on the scale of the job.

\begin{figure}[!t]
    \centering
    \includegraphics[width=0.7\linewidth]{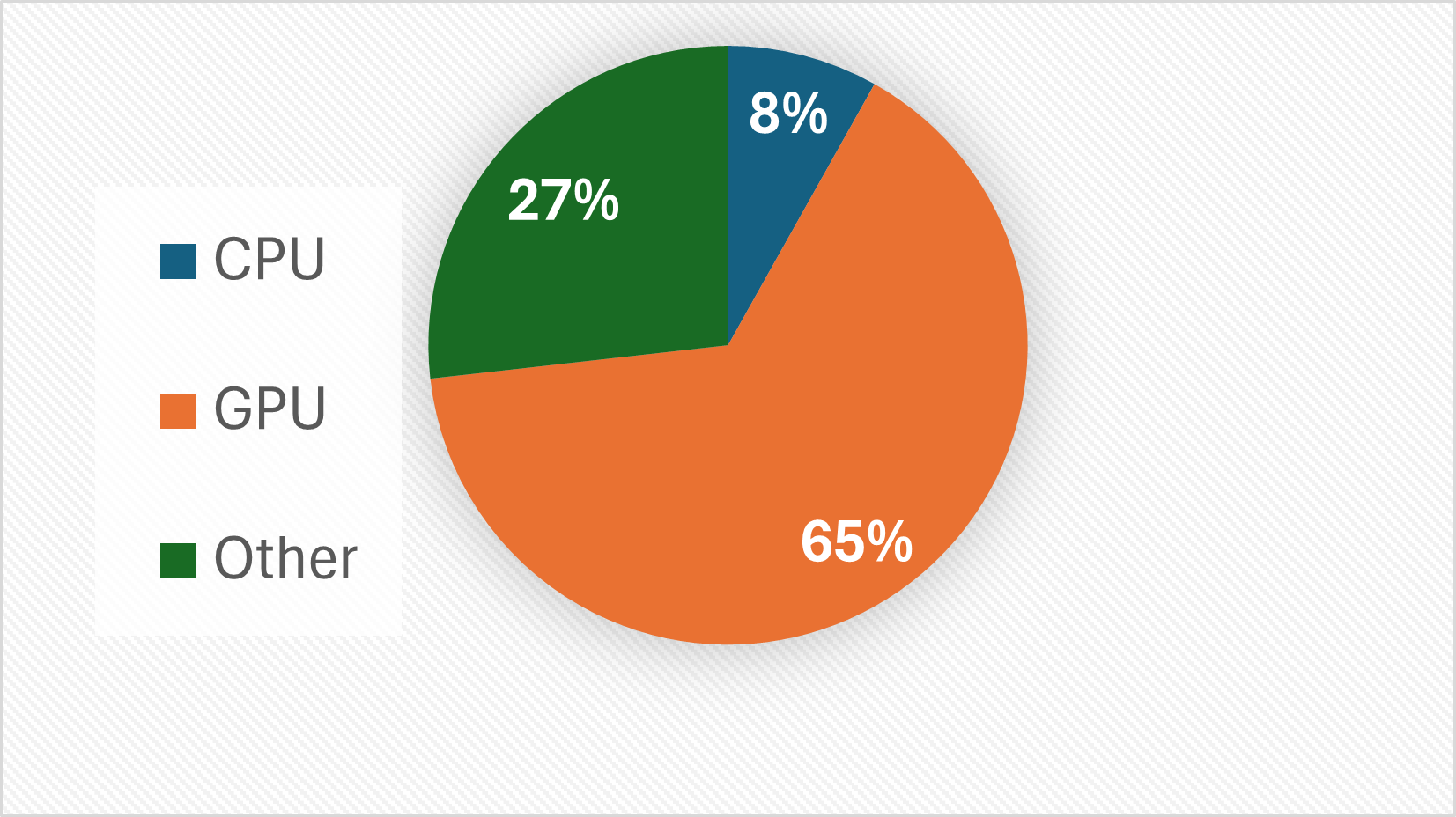}
    \caption{GB200 server power breakdown.}
    \label{fig:powerperc}
\end{figure}

At the server-level, GPUs contribute more than 50\% of the provisioned power, as shown in Figure~\ref{fig:powerperc}. Therefore, these cyclical changes in utilization across thousands of tightly synchronized GPUs manifest as large-amplitude power swings at the rack, row, and potentially even the datacenter power feed levels. The amplitude of these swings grows proportionally with the number of GPUs participating in the job and their peak per-GPU power. In extreme cases, aggregate power consumption can oscillate by tens of megawatts within a single datacenter~\cite{Supermicro_xAI_Colossus_2024}. Such high-magnitude variability can strain power distribution units, affect upstream transformers, and introduce harmonic frequencies that may interfere with the broader utility grid.

In summary, while modern GPU clusters continue to deliver exceptional FLOPS (floating point operations per second) performance for large-scale AI training, the inherent synchronicity of deep learning workloads leads to pronounced swings in power draw. These swings pose new challenges in utility and grid-level stability, motivating the need for systemic solutions that address variability at the software, hardware, and infrastructure layers. 

\subsection{Challenges to generation systems}
The higher-order risk for this  emerges when these cyclical load fluctuations align with or excite torsional resonances in upstream turbine-generator powertrains~\cite{GE_TorsionalDynamics_2013, EPRI_TorsionalInteraction_2006}. As documented in analyses of large steam turbine rotors, torsional vibration occurs at particular natural frequencies, and an external disturbance at or near these frequencies can induce high-amplitude oscillations in the rotor shafts. Prolonged resonance, even if partial, risks mechanical fatigue or shaft failure, particularly in large 2-pole and 4-pole turbine designs.

\subsection{Challenges at utility grid systems}
AI workload frequencies fall into the sub-synchronous regime. These can excite resonant modes in transmission networks potentially leading to sub-synchronous resonance (SSR)~\cite{Wang2017_SSR_SSO} or inter-area oscillations.
Furthermore, rapid, cyclical power swings can lead to voltage flicker visible on lighting systems and frequency modulation that impacts grid frequency regulation, particularly in isolated or constrained regions.

\begin{figure*}[!t]
    \centering
    \includegraphics[width=0.8\linewidth]{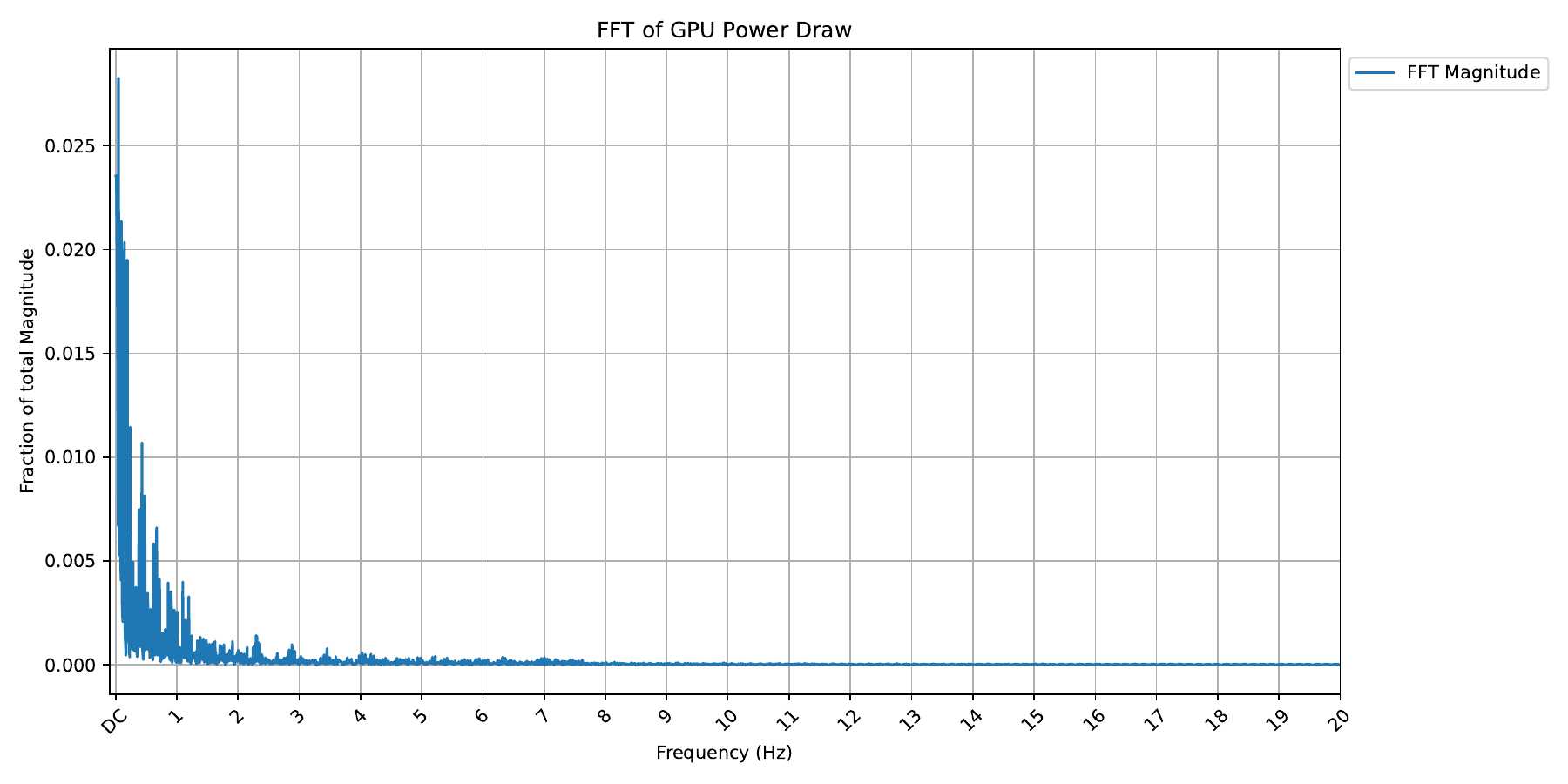}
    \caption{Frequency components of the power waveform shown in Figure 1.}
    \label{fig:fftexample}
\end{figure*}

This is further complicated by the size of the training clusters: as GPU counts grow, the aggregate load swing amplitude at these critical frequencies increases (Figure~\ref{fig:fftexample}), magnifying the potential resonance effects in any interconnected turbine or generator.
To eliminate the possibility that shaft damage or other less severe effects (e.g., breakers opening resulting in islands) occur, prevention of the excitation of these critical frequencies is required.

\section{Specifications and Requirements}
\label{specs}
We present the utility-level specifications and additional recommended requirements for effective mitigation strategies.

\begin{figure*}[tbh]
    \centering
    \includegraphics[width=0.75\linewidth]{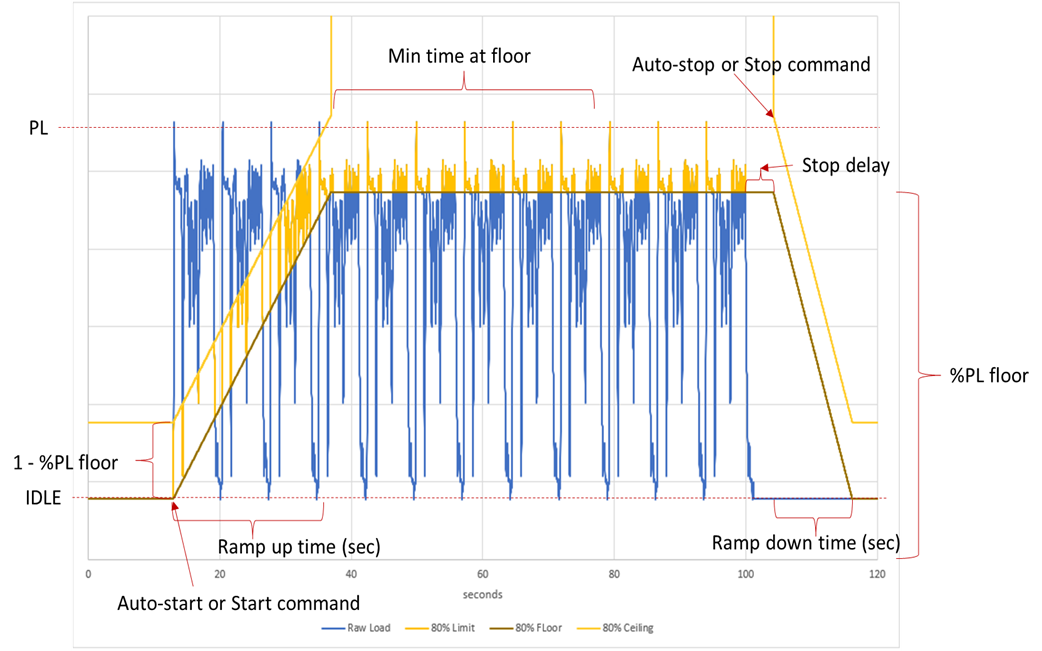}
    \caption{Specification for power stabilization in time domain. PL denotes the device TDP here, and floor denotes MPF. Ramp-up and ramp-down times are calculated from the programmed ramp rates. Power between MPF and ceiling is allowed as a dynamic power range.}
    \label{fig:spec}
\end{figure*}

\subsection{Utility-level specification}
\label{sec:specs}
The specification can vary from one utility to the
other. 
Utility-level specification can be divided into \textit{time-domain spec} and \textit{frequency-domain spec}.

1.\textbf{ Time-domain spec.}
Utility operators may impose \textit{time-domain constraints} on how quickly a load may change its power draw. These constraints include:

\begin{itemize}
    \item \textbf{Ramp-Up Rate}: The maximum permitted rate of increase in power demand, typically expressed in megawatts per second (MW/s).
    \item \textbf{Ramp-Down Rate}: The maximum permitted decrease in power consumption over time.
    \item \textbf{Dynamic Power Range}: The allowed short-term deviation in power draw before ramp constraints are triggered.
\end{itemize}

Figure~\ref{fig:spec} illustrates the time-domain spec with a power waveform.

Time-domain constraints ensure the grid can respond effectively to load changes without triggering oscillations or frequency disturbances.
In practice, utilities monitor planned versus actual power usage using 5- to 15-minute scheduling intervals. Deviations beyond allowed margins may result in financial penalties or curtailment notices. These planning constraints are vital for maintaining reliability in modern power systems, especially in real-time operations and day-ahead markets.

The \textit{dynamic power range} specification is particularly relevant for fast-changing workloads. It defines how much instantaneous fluctuation in power draw is acceptable, usually over sub-second intervals. These tolerances are often informed by grid standards such as IEC 61000-3-3~\cite{IEC_61000_3_3_2013}, which sets thresholds for allowable voltage flicker and short-term harmonic disturbances.

2. \textbf{Frequency-domain spec.}
In addition to ramp limits, utility providers need \textit{frequency-domain specifications} to prevent resonance with power system components. AI workloads, due to their periodic, synchronized nature, can emit power oscillations that align with grid or generator resonant frequencies, causing sub-synchronous resonance (SSR), voltage flicker, or equipment stress.

A typical frequency-domain spec includes:
\begin{itemize}
    \item A \textbf{critical frequency range}, e.g., $0.1-20$ Hz.
    \item A \textbf{maximum allowed spectral magnitude}, e.g. capped at $20\%$ of total harmonic energy within that range.
\end{itemize}

AI workload power traces like those in Figure~\ref{fig:powerbg} show FFT energy concentrated between 0.2--3\,Hz (Figure~\ref{fig:fftexample})—close to known resonant modes of turbine-generator shafts and long transmission lines. As workload behavior evolves, the emitted frequency can shift, so specs must cover a dynamic range.

\subsection{Understanding the frequency-domain spec}
The primary issue unique to a periodic large synchronous load
is the excitation of resonant frequencies in the power generation
and delivery network. Resonant frequencies exist from approximately
0.16Hz to greater than 60Hz in various components of the system:

1. $<$1Hz – Resonance in this range is present due to the naturally
occurring characteristics of long transmission lines connecting
portions of the grid which are independently strong.
A 2019 NERC study~\cite{NERC_Interconnection_Oscillation_Analysis_2019} of interconnection-wide oscillatory behavior observed that most dominant modes are highly damped, though damping varies based on system topology and configuration.
The damping ratio refers to the ability of the
system to stop oscillating after a step excitation and values much
greater than 1 are desirable as they indicate greater speed
at which the resulting oscillations are reduced.
In the case of periodic load variations, there is a danger of instability
despite the damping ratios being greater than one.  An incident in January 2019
caused by an unstable combined cycle unit in Florida quickly grew in magnitude
to a somewhat stable point and not until the unit was taken out of service did the oscillations cease~\cite{nerc2019disturbance}. 
The magnitude of the driving source was approximately 200MW. 
The potential magnitude of the synchronous portion of the load for large training jobs can 
be much greater.
While the oscillations produced during the event in 2019 did not
cause serious disruptions, a much larger distributed load could have greater consequences.

2. $~$1Hz to $~$2.5Hz – In this range, oscillations between closely coupled sources are possible, for instance, units in the same plant or plant to plant in close proximity.  

3. $~$7 Hz to $>$100 Hz – This is the range occupied by shaft torsional
critical frequencies as discussed earlier.  These are the result of one
of more masses on a turbine-generator set oscillating against the other
masses on the shaft.  In a large steam turbine generator, there are many stages.
Typically, the high-pressure stage, the re-heat stage and then one or more low-pressure stages.
Each of these stages is coupled by a length of shaft with the final connection existing between the
last low-pressure section and the generator where the torque from the entire output of all turbine
sections is transmitted.

\subsection{Additional requirements}
Successful mitigation of the large, synchronous power swings must satisfy four additional requirements:

\begin{itemize}

\item \textbf{Ability to meet different utility specifications}.
Power regulations and reliability constraints vary across utilities and geographic regions. Therefore, any proposed intervention must be adaptable to diverse power quality standards and address the range of critical frequencies identified by different transmission and distribution operators.

\item \textbf{Minimal performance loss.}
AI training jobs run for extended periods at substantial computational expense. Any load-shaping or smoothing mechanism should introduce negligible degradation to training throughput or convergence times. Solutions that impose undue latency or stall critical training steps risk driving up operational costs, resources, and undermining the effectiveness of large-scale model development.

\item \textbf{Minimal wasted energy.}
The ideal solution should reduce or eliminate power variability with little to no net increase in energy consumption, preserving both cost-effectiveness and sustainability goals.

\item \textbf{Control EDP.}
Datacenter GPUs allow power overshoots at shorter time-scales (50ms), called \textit{electrical design power (EDP)} peaks (or, EDPp)~\cite{NVIDIA_GB200_NVL_PowerThermals_2025}, while still maintaining the TDP at a granularity of 1 second.
These peaks can be seen in Figure~\ref{fig:mpfgb200} each time the workload spikes up.
Although usually the systems are designed such that the EDP peaks should not be visible beyond the rack power supply units (PSUs), depending on the system design, this may change.
If the EDP peaks are visible at the PDU and utility level, the EDP might need to be set to a lower value to ensure compliance with the utility spec.

\end{itemize}
\section{Mitigation Strategies}

The critical frequencies, and the magnitude acceptable at a critical frequency, together represent a specifications from the utility.
The specification can vary between utilities.
Solutions in this space should be able to meet even the most stringent specifications. 

\subsection{Software-only mitigation}

A purely software-based approach to mitigating power swings from the primary workload focuses on dynamically introducing power-hungry secondary workloads such as GEMM kernels, whenever the GPU activity level or power falls below a designated threshold. By doing so, the system can sustain a more uniform power draw across the compute and communication phases of each training iteration, thereby reducing the amplitude of power fluctuations visible to the datacenter and utility infrastructure.

\textbf{Secondary workload.}
This power-hungry secondary workload could either be a lower-priority, useful job, or an artificial workload.
The challenge with a useful job is that the job's state would need to be saved and restored, causing additional delay in power smoothing, and performance impact to the primary workload.
An artificial workload would avoid such challenges, with the downside of wasted energy.

\textbf{Monitoring.}
An important note is that the secondary workload cannot be deterministically called within a communication library. The core of the problem is that the GPU power drop is the result of several compute kernels ending, rather than a communication kernel starting. Therefore, instead of a compiler-based solution, we prefer a software mechanism that uses real-time, fine-grained power and activity monitoring of the GPUs.
Today, NVIDIA datacenter GPUs have capability to expose instantaneous or averaged in-band power and activity readings at a minimum of 1-100ms latency, depending on the acceptable reliability of the counters.
The reliable 100ms counters are too slow for a use-case where we would want to detect power swings at 20Hz for instance, needing us to inject a secondary workload every 50 ms.
Therefore, this solution will have to be built upon faster telemetry sources.

\textbf{Secondary workload start and stop conditions.}
The secondary workload is started as soon as the block activity (and thus power draw) drops below a preset level, preventing a drastic power dip. As soon as the utilization for the primary workload ramps up again, the secondary workload needs to back off. However, since there are no activity or power counters per process today, the secondary workload will need to periodically back off and read the activity counters.

The ramp-up and ramp-down requirement is easy to meet in a software-mitigation by either stepping up the load synchronously, or staggering the load ramp-up across all the participating GPUs. 

\textbf{Firefly.}
With these insights, we built a solution, named \textbf{\textit{Firefly}}, to mitigate the power swings in software.
We used NVIDIA's Multi-Process Service (MPS), which allows multiple CUDA processes to share a single GPU context, with the provisioning to carve up resources as needed.
For monitoring we used block activity counters from the GPU, and as a secondary workload, we used a series of matrix multiplications scaled as needed.
Firefly was able to increase the power utilization all the way up to 100\% of the TDP.

\textbf{Challenges.}
Although Firefly does not require any additional hardware features or infrastructure for mitigation, it does present a few challenges.
The first one is related to performance overheads.
The secondary workload needs some memory and compute resources of its own. Additionally, the back-off based mechanism for stopping the secondary workload leads to performance loss for the primary workload.
With NVIDIA's MPS, we were able to bring down the performance overhead to the primary workload down to $<5\%$. However, there was a considerable amount of CPU cores and host-device bandwidth dedicated for processing the GPU power data continuously at a 1 ms granularity. As training jobs become more heterogeneous, requiring more CPU work, this can become a challenge too, making the software solution very expensive from a resource standpoint.

Second, since the GPUs are used in direct-device access mode by most virtual machine offerings from large cloud providers today, this mitigation requires close collaboration between the end customer and the cloud provider, which might be difficult in certain settings.
The solution may even need tuning as the primary workload changes, to ensure that the primary power patterns are well hidden.

Third challenge is the reliability of this solution. In particular, the MPS feature ties the failure domain of the primary and secondary workloads at the GPU scale, since they share the same GPU context. A fault in one workload can propagate to the other, making the setup more error-prone at scale. Improving reliability will require software enhancements across multiple layers of the stack.

And finally, an artificial secondary workload, if not performing real work, wastes energy.
\begin{figure}[tbh]
    \centering
    \includegraphics[width=1\linewidth]{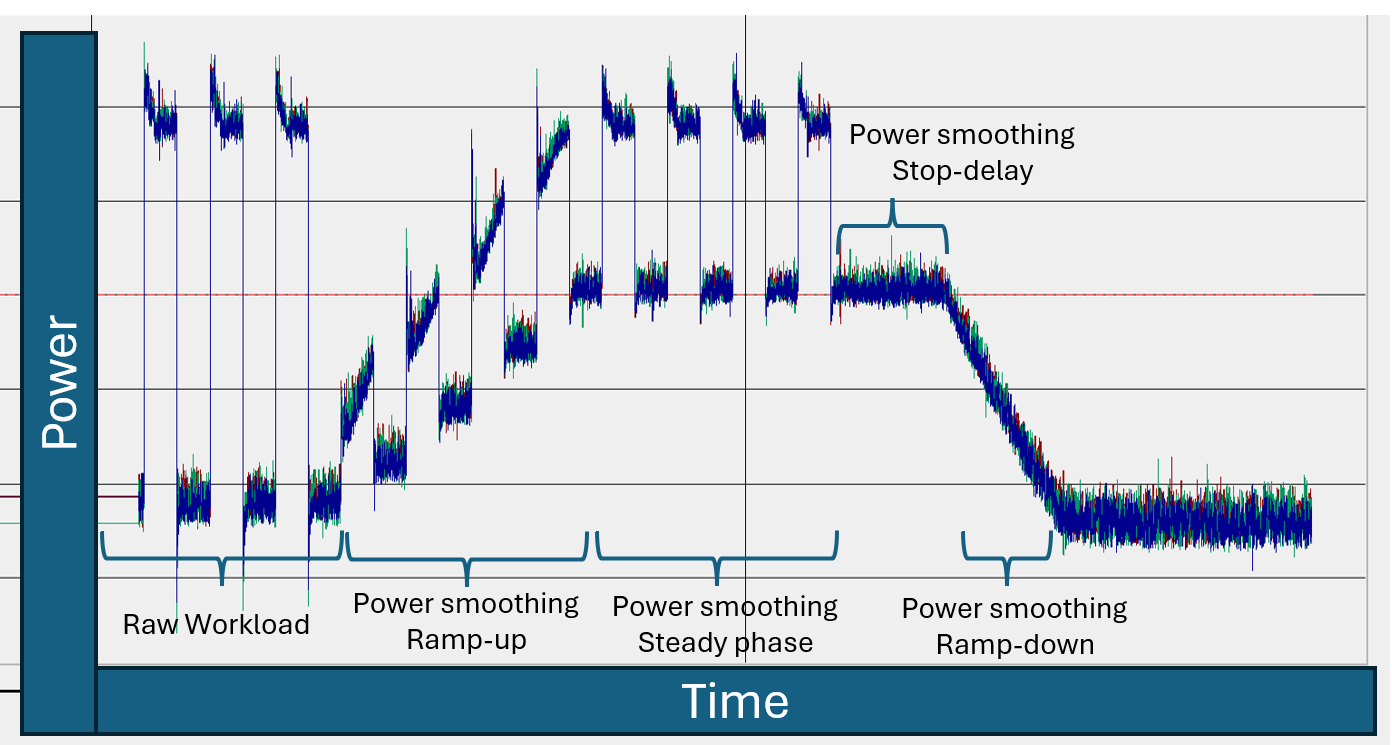}
    \caption{GB200 Power smoothing results with a square-wave microbenchmark.}
    \label{fig:mpfgb200}
\end{figure}

\textbf{Potential optimizations.}

1. \textit{Separate failure domains:} If the primary and secondary jobs could be run in a way that the failure of secondary job does not lead to a crash in the primary workload, the at-scale reliability of this solution would benefit a lot.

2.\textit{ Adaptive throttling of the secondary workload:} Instead of being completely dependent on monitoring, the software could try predictive modeling of forthcoming communication/computation phases. This could help reduce the performance impact of this solution.

3.\textit{ Priority scheduling:} Faster, priority-based scheduling mechanisms in the GPU could further avoid the need to self-preempt the secondary workload, reducing the interference for the primary workload.

4.\textit{ Software solution in the firmware:} The hardware vendor could provide the software solution in a way that is tightly coupled with the scheduler and monitoring. This would reduce the overheads, and the dependency on the end customer.

\begin{figure*}[tbh]
    \centering
    \includegraphics[width=1\linewidth]{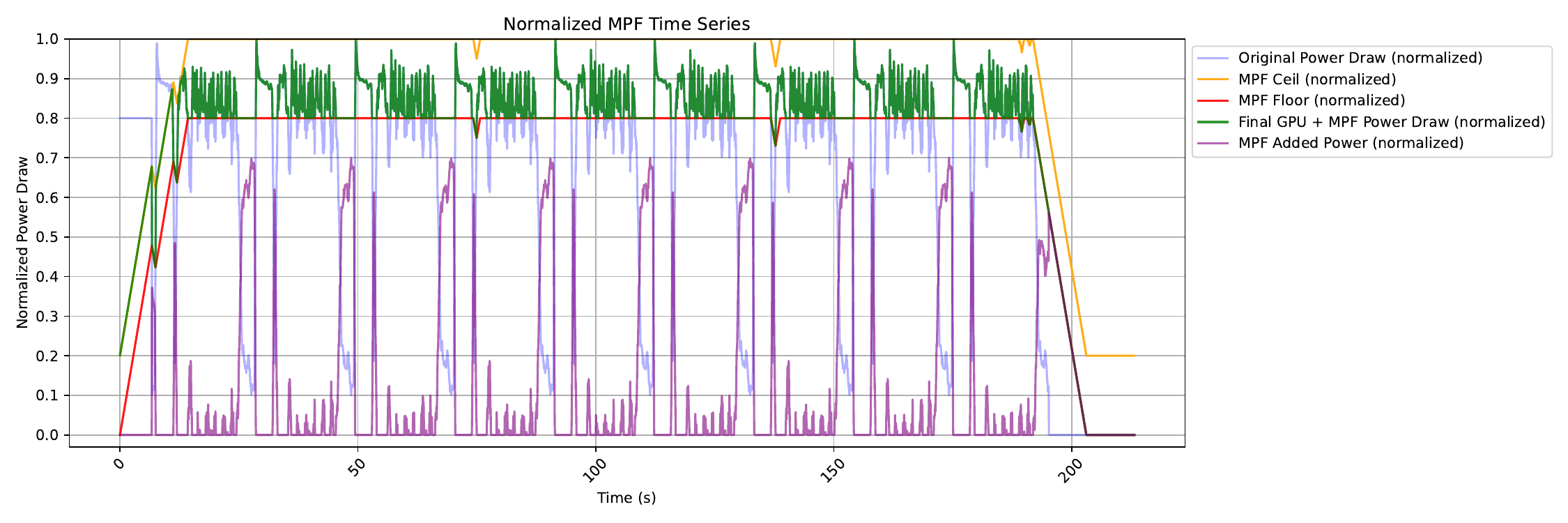}
    \caption{Power smoothing to the minimum power floor (MPF) simulated on the training waveform from Figure~\ref{fig:powerbg}.}
    \label{fig:mpf}
\end{figure*}

Overall, software-based smoothing offers a flexible and relatively quick-to-deploy solution, providing immediate relief from large power swings without hardware modifications. However, it requires a careful trade-off between limiting performance impact and minimizing wasted energy, thus underscoring the importance of precise telemetry, well-designed fallback logic, and ongoing calibration as the workloads scale.

\subsection{GPU power smoothing}
To avoid the performance loss, secondary workload tuning with primary workload changes, and the close collaboration requirements in cloud provider settings, we discuss a possible GPU-level solution. The GPU power smoothing feature, as introduced in NVIDIA GB200, allows the developers in-band, and the cloud provider out-of-band, to program a preset profile to each GPU. 
The profile includes the following notable settings:

1. \textbf{Ramp-up rate and ramp-down rate:}
These settings can be programmed to directly meet the requirements of utility's time domain spec. These can be programmed in a watts per second format. 

2. \textbf{Minimum Power Floor (MPF):}
This setting allows the user to set the floor for the power utilization by the GPU during stable periods of execution.
In combination with the maximum allowed power (thermal design power) of the GPU, this allows us to meet the dynamic power range of the utility's spec, ensuring that the power changes in short time-scales are within the spec. 

3. \textbf{Stop delay:}
This setting specifies how long the GPU should stay at the minimum power floor without any real workload activity before ramping down. This stop delay presents a trade-off between performance impact and energy burn, while meeting the spec. 

Since this feature implementation uses hardware counters and the GPU's power controller, the power floor engage and disengage latencies can be very small, allowing us to meet the frequency spec from the utility.
Note that a high MPF value will maximize the energy burn.
On the other hand, the dynamic power range specification leads to least performance impact at a high MPF.

Figure~\ref{fig:mpfgb200} shows this feature on a GB200 with a square-wave power micro-benchmark with high power utilization across all the GPUs.
The power floor was set to $65\%$ of the TDP of the GPU. 
The figure shows the ramp-up, steady phase, stop delay, and ramp-down phases of the run. Note that the workload stays the same throughout, until the stop delay phase, where the workload has no activity.

Since the micro-benchmark shown in Figure~\ref{fig:mpfgb200} is not representative of the true ups and downs in the power waveform during a training workload, we use Microsoft's in-house power simulator called StratoSim, to show the impact of power smoothing on the real waveform from Figure~\ref{fig:powerbg}. Figure~\ref{fig:mpf} shows the results.
The floor was set to $90\%$ of the TDP for this simulation. 
At such a high setting, for the power waveform in Figure~\ref{fig:powerbg}, we note a total energy overhead of $10.5\%$ due to the power smoothing feature.

\textbf{Challenges.}
Similar to the software solution, the additional energy burned with the GPU power smoothing is the main downside of this solution.
The GB200 solution includes a built-in lifetime counter because its durability is limited. The actual lifetime depends on how frequently the feature is used and the extra energy it consumes. While the expected lifespan is reasonable compared to the typical 5-year GPU lifetime, real usage in training jobs over the coming years will provide a more accurate picture of its endurance.
Furthermore, since the current solution in GB200 only allows a maximum MPF of 90\% of TDP, and a minimum EDP of 1.1$\times$ of TDP, we are left with at least 20\% of the TDP as the dynamic range (Section~\ref{sec:specs}) of the power fluctuations at the GPU level. This creates challenges in meeting tighter specs.
Let us say for instance, the utility company requires us to meet a 10MW dynamic range, and we are running a job using 100MW of power. This requires a dynamic power range of 10\% being allowed, which cannot be met with GPU power smoothing today.

\begin{figure*}[tbh]
    \centering
    \includegraphics[width=1\linewidth]{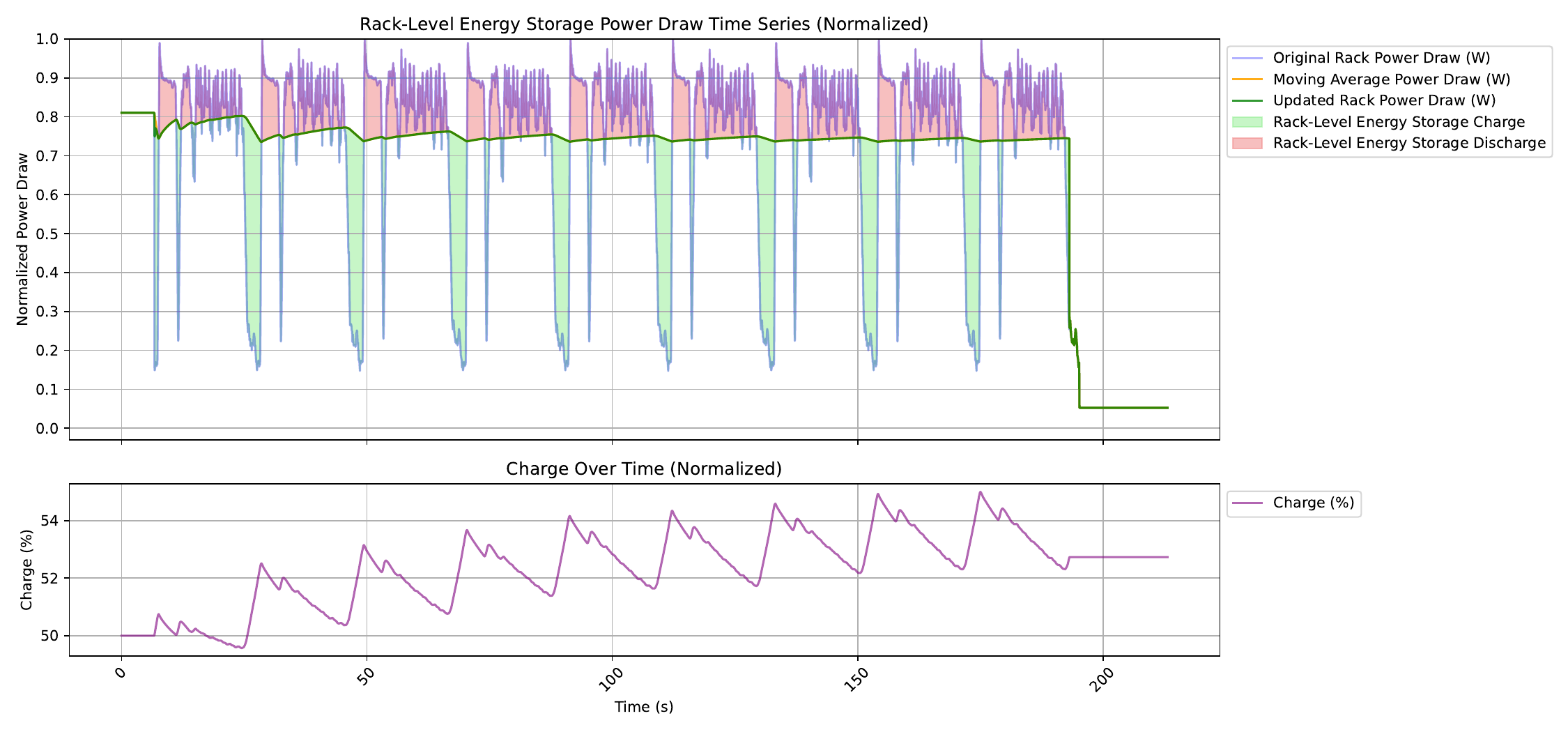}
    \caption{Energy-storage solution simulated on the power waveform from Figure~\ref{fig:powerbg}.}
    \label{fig:pcu}
\end{figure*}

\subsection{Energy-storage solution}
The best-case solution to the training power-stabilization challenge is an energy-storage solution that

1. Can directly measure the load,

2. Has enough capacitance to support the workload,

3. Can meet the sudden rise/drop needs in power, and

4. Can switch modes between charging and discharging quickly.

An energy-storage solution does not waste energy to solve the power stabilization problem.
Instead, it can potentially even reduce peak power needs in training, by using the low-power communication phases to charge, and release energy during high-power computation phases. Figure~\ref{fig:pcu} shows a simulated example of how this solution could work. We show the battery charge along with the final waveform.

\textbf{\textit{Placement level.}}
The main design question concerns the placement of the energy-storage solution.
It could be added at the server-level, rack-level, row-level, or colo/datacenter-level.
The higher up in the hierarchy we add the energy storage, more devices like UPSes and PDUs become exposed to the power and voltage perturbations.
Although a higher hierarchy level can theoretically offer more power demand multiplexing from the servers, since we are concerned about large synchronous training jobs that have identical power demands across all participating servers, this is not a factor that affects us.
A more distributed energy-storage also allows for a more relaxed reliability metric, since the failure domain of a rack-level should not have a big effect on the datacenter-wide power waveform.
Furthermore, the AC-DC converters are present at the rack-level already, making that an optimal place for a DC block energy storage.
Therefore, a rack-level energy storage emerges as the best option.
\begin{table*}[t]
\centering
\small
\renewcommand{\arraystretch}{1.3}
\begin{tabular}{lccccccc}
\hline
\textbf{Solution} & \textbf{Reliability} & \textbf{Performance} & \textbf{Energy} & \textbf{Cost} & \textbf{Ability to meet}  & \textbf{Dependency on the} & \textbf{Lifetime}
\\& & & & & \textbf{tightest spec} & \textbf{developer}\\
\hline
Software-only mitigation 
& Medium 
& Medium 
& High 
& Medium 
& High 
& High
& High\\

GPU power smoothing 
& High 
& Medium 
& High 
& Low 
& Medium 
& Medium
& Medium\\

Rack-level energy storage 
& High 
& High 
& Low 
& High 
& High 
& Low
& High\\
\hline
\end{tabular}
\caption{Summary of various proposed solutions. For energy, cost, and dependency on the developer, lower is better.}
\label{tab:mitigation-summary}
\end{table*}

\textbf{Challenges.}
Given the range of frequencies that need to be handled, one of the main challenges that emerges is an energy storage solution that can meet demands across this spectrum.
Higher frequencies are easier to filter out, compared to the lower frequencies. 
Additionally, ramp-up and ramp-down can require very large capacitance from the energy storage.
Such large capacitance would be very expensive from cost, rack-level space, and embodied carbon perspectives.
Given that these events happen rarely, compared to the rest of the workload run, designing enough capacity for this does not necessarily pay off.

\subsection{Putting a solution together.}

The three solutions we discussed have various pros and cons.
We summarize these in Table~\ref{tab:mitigation-summary}.

The software and hardware solutions are readily available with the latest hardware, and both have similar energy overheads. The hardware solution is much more reliable at scale today, and does not lead to resource overheads unlike the software-solution.
However, in regions where utilities enforce stricter time-domain specifications (e.g., dynamic range power), hardware alone may not be sufficient due to the 90\% limit on MPF; in such cases, a Firefly-like solution can be used in combination to meet the requirements.
The energy storage solution requires additional hardware deployment, leading to higher cost and embodied carbon in lieu of lower training energy. Using energy storage alone would require high capacitance to meet the ramp-up and ramp-down phase specs.

Therefore, we propose a combination of the solutions: GPU level power smoothing using software/hardware solutions, and a rack-level energy storage.
The GPU-level solution can be used to meet the ramp-up, ramp-down specs, and any corner case scenarios where the energy storage runs out of capacity.
Such a combination of solutions is optimal from wasted energy, cost, and space perspectives.
However, it does require a co-design of the solutions, such that the rack-level energy storage and the GPU should be able to communicate regarding the state of charge.

For even larger AI training deployments in the future, long storage BESS (battery energy storage system) should also be considered. 
In general a combined approach of solutions closer to the rack (the mitigation strategies we discussed in detail), supplemented with battery storage systems at larger scale would help alleviate the power swing challenges.

\subsection{Fast telemetry-based backstop}

While proactive smoothing strategies can mitigate most power fluctuations, large-scale AI training jobs can still occasionally excite critical sub-synchronous frequencies. To safeguard against this, a fast telemetry-based backstop system is necessary. This system continuously monitors power waveforms across the datacenter, looking for early signs of instability or resonance that may not be addressed in time by primary mitigation techniques.

By leveraging fine-grained, low-latency telemetry and real-time spectral analysis (e.g., FFT bin monitoring), the system can identify the emergence of problematic frequencies and initiate tiered responses. Initial interventions might include soft throttling or load shaping; if ineffective, more aggressive responses can follow—such as circuit-level power shedding or coordinated disconnects—executed in collaboration with site-specific infrastructure logic.

\section{Call to Action}

As the scale and complexity of AI training workloads continue to grow, power variability and grid impact will only become more severe. Addressing this challenge requires proactive collaboration across software, hardware, infrastructure, and utility domains. We outline three critical calls to action:

1. AI Framework and System Designers: Explore less synchronous, more power-aware training algorithms that reduce large-scale power swings without compromising convergence. This includes asynchronous training techniques, staggered scheduling, and overlap of compute and communication.

2. Utility Providers and Grid Operators: Share resonance and ramp specifications openly and establish standardized communication pathways with datacenter operators. This coordination is essential to ensure safe grid operation and avoid unplanned outages or equipment degradation.

3. Industry Collaboration: Support and participate in pre-competitive, open forums—such as the Open Compute Project (OCP)—to establish interoperable standards for telemetry, load signaling, and sub-synchronous oscillation mitigation. It is extremely difficult for a single customer, vendor, or hyperscaler to solve this problem in isolation.

\section{Conclusion}
Power stabilization is emerging as a critical bottleneck in the continued scaling of AI training workloads.
In this paper, we detailed the impact such workloads can lead to, and gave examples of the specifications needed for mitigation.
We have presented a cross-stack approach that combines software-based smoothing, GPU-level controls, and rack-level energy storage, backed by real-world measurements and simulation. These techniques offer practical and immediate relief for today's deployments.

However, this work is just the beginning. Ensuring that AI infrastructure remains both performant and grid-safe will require sustained collaboration across research, industry, and utilities. We urge the community to come together—through venues such as the Open Compute Project (OCP) and beyond—to drive forward shared standards, validation frameworks, and architectural best practices. Together, we can design for a future where AI training is not only powerful, but also power-aware.

\bibliographystyle{plain}
\bibliography{bib}

\begin{thebibliography}{10}

\bibitem{IEC_61000_3_3_2013}
Iec 61000-3-3:2013 — electromagnetic compatibility (emc), 2013.
\newblock Standard for voltage flicker and power swing limitations.

\bibitem{brown2020languagemodelsfewshotlearners}
Tom~B. Brown, Benjamin Mann, Nick Ryder, Melanie Subbiah, Jared Kaplan, Prafulla Dhariwal, Arvind Neelakantan, Pranav Shyam, Girish Sastry, Amanda Askell, Sandhini Agarwal, Ariel Herbert-Voss, Gretchen Krueger, Tom Henighan, Rewon Child, Aditya Ramesh, Daniel~M. Ziegler, Jeffrey Wu, Clemens Winter, Christopher Hesse, Mark Chen, Eric Sigler, Mateusz Litwin, Scott Gray, Benjamin Chess, Jack Clark, Christopher Berner, Sam McCandlish, Alec Radford, Ilya Sutskever, and Dario Amodei.
\newblock Language models are few-shot learners, 2020.

\bibitem{chowdhery2022palm}
Aakanksha Chowdhery, Sharan Narang, ..., Jason Wei, and [and many~others] ...~Petraﬁkowski.
\newblock Palm: Scaling language modeling with pathways.
\newblock In {\em JMLR Workshop and Conference Proceedings}, 2023.
\newblock 540‑billion‑parameter model.

\bibitem{GE_TorsionalDynamics_2013}
General~Electric Company.
\newblock Torsional dynamics: Large 2‑pole and 4‑pole steam turbine powertrains.
\newblock Technical report (ger‑4724), GE Power \& Water, 2013.
\newblock Based on EPRI 1011679, Electric Power Research Institute, 2005.

\bibitem{deepseek2024v3}
DeepSeek‑AI, Aixin Liu, Bei Feng, Bing Xue, ..., and many others.
\newblock Deepseek‑v3 technical report.
\newblock Technical report, DeepSeek‑AI / CoRR, December 2024.
\newblock Mixture‑of‑Experts language model with 671 B parameters (37 B activated per token).

\bibitem{EPRI_TorsionalInteraction_2006}
{Electric Power Research Institute}.
\newblock Torsional interaction between electrical network phenomena and turbine‑generator shafts: Plant vulnerability.
\newblock Technical Report 1013460, EPRI, Palo Alto, CA, 2006.

\bibitem{krizhevsky2012imagenet}
Alex Krizhevsky, Ilya Sutskever, and Geoffrey~E. Hinton.
\newblock Imagenet classification with deep convolutional neural networks.
\newblock In {\em Advances in Neural Information Processing Systems 25}, pages 1097--1105. Curran Associates, Inc., 2012.

\bibitem{lian2018asynchronous}
Xiangru Lian, Wei Zhang, Ce~Zhang, and Ji~Liu.
\newblock Asynchronous decentralized parallel stochastic gradient descent.
\newblock In Jennifer~G. Dy and Andreas Krause, editors, {\em Proceedings of the 35th International Conference on Machine Learning (ICML)}, volume~80 of {\em Proceedings of Machine Learning Research}, pages 3049--3058. PMLR, 2018.

\bibitem{meta2025aireliability}
{Meta Engineering}.
\newblock How meta keeps its ai hardware reliable.
\newblock Engineering blog, July 2025.
\newblock Accessed: 2025‑08‑06.

\bibitem{abdin2024phi3}
{Microsoft Azure AI Team}.
\newblock Phi‑3: A highly capable small language model locally on your phone.
\newblock Technical report, Microsoft, April 2024.
\newblock Introduced in Microsoft Azure AI blog; technical report available on arXiv.

\bibitem{nerc2019disturbance}
{North American Electric Reliability Corporation (NERC)}.
\newblock Disturbance monitoring and analysis of oscillatory events.
\newblock \url{https://www.nerc.com}, 2019.
\newblock [Online; accessed 2025-08-06].

\bibitem{NERC_Interconnection_Oscillation_Analysis_2019}
{North American Electric Reliability Corporation (NERC)}.
\newblock Interconnection oscillation analysis.
\newblock Reliability assessment technical report, North American Electric Reliability Corporation, July 2019.
\newblock Report published July 2019; includes analysis of inter‑area oscillations, notable events (e.g., Alberta separation 2000, WECC 2005, EI 2016), and modal characteristics of the Eastern, Western, and ERCOT interconnections.

\bibitem{NVIDIA_GB200_NVL_PowerThermals_2025}
NVIDIA.
\newblock {\em NVIDIA GB200 NVL Multi-Node Tuning Guide — Power and Thermals}.
\newblock NVIDIA, April 2025.
\newblock Provides GPU power and thermal management tuning for data center systems.

\bibitem{nccl2025}
{NVIDIA Corporation}.
\newblock Nvidia collective communications library (nccl).
\newblock \url{https://developer.nvidia.com/nccl}, 2025.
\newblock Accessed: 2025‑08‑06.

\bibitem{OpenAI_ScalingK8s_2021}
OpenAI.
\newblock Scaling kubernetes to 7{,}500 nodes.
\newblock \url{https://openai.com/index/scaling-kubernetes-to-7500-nodes/}, January 2021.

\bibitem{OpenAI_TechniquesTraining_2022}
OpenAI.
\newblock Techniques for training large neural networks.
\newblock \url{https://openai.com/index/techniques-for-training-large-neural-networks/}, June 2022.

\bibitem{patel2024llmpower}
Pratyush Patel, Esha Choukse, Chaojie Zhang, {\'I\~n}igo Goiri, Brijesh Warrier, Nithish Mahalingam, and Ricardo Bianchini.
\newblock Characterizing power management opportunities for llms in the cloud.
\newblock In {\em Proceedings of the 29th ACM International Conference on Architectural Support for Programming Languages and Operating Systems (ASPLOS)}, volume~3, pages 207--222, La Jolla, CA, USA, 2024. Association for Computing Machinery.

\bibitem{pilz2025trendsaisupercomputers}
Konstantin~F. Pilz, James Sanders, Robi Rahman, and Lennart Heim.
\newblock Trends in ai supercomputers, 2025.

\bibitem{shoeybi2019megatronlm}
Mohammad Shoeybi, Mostofa Patwary, Raul Puri, Patrick LeGresley, Jared Casper, and Bryan Catanzaro.
\newblock Megatron‑lm: Training multi‑billion parameter language models using model parallelism.
\newblock arXiv preprint arXiv:1909.08053, 2019.
\newblock \url{http://arxiv.org/abs/1909.08053}.

\bibitem{Supermicro_xAI_Colossus_2024}
{Super Micro Computer, Inc.}
\newblock Inside the 100k gpu xai colossus cluster that supermicro helped build for elon musk.
\newblock Case study / success story, Super Micro Computer, Inc., December 2024.
\newblock \url{https://www.supermicro.com/CaseStudies/Success_Story_xAI_Colossus_Cluster.pdf}.

\bibitem{touvron_llama_2023}
Hugo Touvron, Thibaut Lavril, ..., and Guillaume Lample.
\newblock Llama: Open and efficient foundation language models.
\newblock arXiv preprint arXiv:2302.13971, 2023.

\bibitem{Wang2017_SSR_SSO}
L.~Wang.
\newblock Review of emerging ssr/sso issues and their classifications.
\newblock {\em Journal of Operational Engineering (JOE)}, 2017.
\newblock Online.

\bibitem{grok1_xai_2024}
{xAI}.
\newblock Open release of grok‑1: A 314b parameter mixture‑of‑experts model.
\newblock Web page, 2024.
\newblock Released March 17, 2024.

\end{thebibliography}

\end{document}